\documentclass[aps,prl,reprint,superscriptaddress]{revtex4-1}
\usepackage{graphicx}
\usepackage{color,amsmath}
\usepackage[normalem]{ulem}
\usepackage{hyperref}
\usepackage{lipsum}
\usepackage{braket}

\usepackage{mathtools}

\newcommand{\vpg}{V_{\rm{pg}}}
\newcommand{\vt}{V_{\rm{t}}}
\newcommand{\vsd}{V_{\rm{sd}}}
\newcommand{\bpl}{B_{\parallel}}
\newcommand{\bperp}{B_{\perp}}

\newcommand{\Ez}{E_{\rm{Z}}}
\newcommand{\zeropi}{0{\text -}\pi}

\newcommand{\Ebs}{E_{\rm{BS}}}

\begin{document}

\title{Zeeman-driven parity transitions in an Andreev quantum dot}

\author{A.~M.~Whiticar}
\altaffiliation[Present address: D-Wave  Systems,  Burnaby,  British  Columbia,  Canada]{}

\author{A.~Fornieri}
\affiliation{Center for Quantum Devices, Niels Bohr Institute, University of Copenhagen and Microsoft Quantum Lab--Copenhagen, Universitetsparken 5, 2100 Copenhagen, Denmark}

\author{A.~Banerjee}
\affiliation{Center for Quantum Devices, Niels Bohr Institute, University of Copenhagen and Microsoft Quantum Lab--Copenhagen, Universitetsparken 5, 2100 Copenhagen, Denmark}

\author{A.~C.~C.~Drachmann}
\affiliation{Center for Quantum Devices, Niels Bohr Institute, University of Copenhagen and Microsoft Quantum Lab--Copenhagen, Universitetsparken 5, 2100 Copenhagen, Denmark}

\author{S.~Gronin}
\affiliation{Department of Physics and Astronomy and Microsoft Quantum Lab--Purdue, Purdue University, West Lafayette, Indiana 47907 USA}
\affiliation{Birck Nanotechnology Center, Purdue University, West Lafayette, Indiana 47907 USA}

\author{G.~C.~Gardner}
\affiliation{Department of Physics and Astronomy and Microsoft Quantum Lab--Purdue, Purdue University, West Lafayette, Indiana 47907 USA}
\affiliation{Birck Nanotechnology Center, Purdue University, West Lafayette, Indiana 47907 USA}
\author{T.~Lindemann}
\affiliation{Department of Physics and Astronomy and Microsoft Quantum Lab--Purdue, Purdue University, West Lafayette, Indiana 47907 USA}
\affiliation{Birck Nanotechnology Center, Purdue University, West Lafayette, Indiana 47907 USA}

\author{M.~J.~Manfra}
\affiliation{Department of Physics and Astronomy and Microsoft Quantum Lab--Purdue, Purdue University, West Lafayette, Indiana 47907 USA}
\affiliation{Birck Nanotechnology Center, Purdue University, West Lafayette, Indiana 47907 USA}
\affiliation{School of Materials Engineering, Purdue University, West Lafayette, Indiana 47907 USA}
\affiliation{School of Electrical and Computer Engineering, Purdue University, West Lafayette, Indiana 47907 USA}

\author{C.~M.~Marcus}
\email[email: ]{marcus@nbi.ku.dk}
\affiliation{Center for Quantum Devices, Niels Bohr Institute, University of Copenhagen and Microsoft Quantum Lab--Copenhagen, Universitetsparken 5, 2100 Copenhagen, Denmark}


\begin{abstract}
The Andreev spectrum of a quantum dot embedded in a hybrid semiconductor-superconductor interferometer can be modulated by electrostatic gating, magnetic flux through the interferometer, and Zeeman splitting from in-plane magnetic field. We demonstrate parity transitions in the embedded quantum dot system, and show that the Zeeman-driven transition is accompanied by a 0-$\pi$ transition in the superconducting phase across the dot. We further demonstrate that flux through the interferometer  modulates both dot parity and 0-$\pi$ transitions. 

\end{abstract}

\maketitle

\section{Introduction}

The interplay of confinement, spin, and superconductivity leads to a rich variety of mesoscopic phenomena~\cite{eschrig2011spin,de2010hybrid,klapwijk2004proximity} that can be investigated in semiconductor-superconductor hybrid materials coupled via the proximity effect \cite{klapwijk2004proximity,ryazanov2001coupling}. Recent advances in epitaxial growth of such hybrids have demonstrated highly transparent heterointerfaces in several material platforms~\cite{krogstrup2015epitaxy,Shabani2016,kjaergaard2017transparent,lutchyn2018majorana}. 

An important application is semiconducting Josephson junctions (JJs), where a semiconducting normal (N) region is bounded by two superconductors (S), giving rise to a spectrum of Andreev bound states (ABSs) in the N region at energies below the gap, $\Delta$, of the superconductors~\cite{beenakker1991universal}. ABS energies $E$ depend on the superconducting phase difference, $\varphi$, across the junction and generate a supercurrent, $I_{\rm s}(\varphi) = -(2e/h)\, dE/d\varphi$, where $e$ is the unit of charge and $h$ is Planck's constant~\cite{yokoyama2014anomalous,Nichele2020}. Semiconducting S-N-S junctions have been used as voltage-controlled transmons (gatemons)  ~\cite{deLange2015,larsen2015semiconductor,casparis2018superconducting,Kringhoej2020,Bargerbos2020} and Andreev qubits~\cite{zazunov2003andreev,janvier2015coherent,Hays2018,Tosi2019}.

A quantum dot (QD) embedded in a Josephson junction (S-QD-S) can result in a competition between superconductivity and spin in a confined system \cite{de2010hybrid,eichler2007even,sand2007kondo,kirvsanskas2015yu,He2020}. The charging energy of a weakly coupled QD typically stabilizes one of two spin states at zero magnetic field, depending on dot occupancy: a spin-zero singlet $\ket{S}$ or a spin-$\frac{1}{2}$ doublet $\ket{D}$\cite{kirvsanskas2015yu,Zitko2016ZBA,Meng2009}. For even dot occupancy, the ground state (GS) is typically a singlet for all coupling strengths; for odd occupancy, the GS is either a $\ket{D}$ state for weak coupling, or a delocalized singlet, where the spin of the dot is hybridized with spins in the leads~\cite{Zitko2016ZBA}. The hybridized odd-parity subgap spectrum corresponds to Yu-Shiba-Rusinov states~\cite{jellinggaard2016tuning,Zitko2016ZBA,Delagrange2016}, and the crossover between even and odd parity ~\cite{kirvsanskas2015yu,Meng2009,vecino2003josephson,rozhkov1999josephson} is marked by a zero-energy crossing~\cite{chang2013tunneling,lee2014spin,jellinggaard2016tuning,Zitko2017}.

Bound-state (BS) excitations at energies $\Ebs$ correspond to the differences between GS and first excited-state (ES) energies~\cite{lee2014spin}. When $\Ebs = 0$, the GS and ES are degenerate and a fermionic parity transition occurs~\cite{vecino2003josephson,Zitko2016ZBA}. The even and odd GS parities can be distinguished by the phase dependence of $\Ebs$. An odd GS leads to a superconducting phase difference of $\pi$ across the JJ, shifting the phase dependence of the subgap excitations by $\pi$ that results in a negative supercurrent~\cite{rozhkov1999josephson,van2006supercurrent}. The GS parity transition from even to odd is commonly referred to as a $\zeropi$ phase transition and can be identified spectroscopically by the observation of zero-bias crossings (ZBCs) and $\pi$-shifted phase dependence of the subgap spectrum via tunneling of single electrons into the junction from a weakly-coupled normal lead~\cite{kirvsanskas2015yu,vecino2003josephson,pillet2010andreev,chang2013tunneling}.


In this Article, we investigate an S-QD-S junction embedded in a superconducting quantum interference device
(SQUID).  The device allowed control of the superconducting phase across the QD by threading magnetic flux through the SQUID loop, while BS energies were simultaneously measured via tunneling spectroscopy into the QD with a third (normal) lead. The system was fabricated from an epitaxial InAs-Al heterostructure patterned using electrostatic gates. We investigate spectra of subgap excitations under the influence of flux and Zeeman field across parity transitions identified by ZBCs. We observe that the subgap spectrum acquires a $\pi$ shifted energy dependence in the odd-parity GS, as expected for a $\zeropi$ phase transition. We find that these transitions are controlled via magnetic field, gate voltage, and superconducting phase difference, which demonstrates precise control of Andreev states in semiconductor Josephson junctions. 

Magnetic-field driven parity transitions in S-QD-S junctions have previously been observed as zeros of a reentrant critical current~\cite{saldana2019charge} or via direct spectroscopy of the QD~\cite{pillet2010andreev,chang2013tunneling}. Related measurements in N-QD-S devices showed gate voltage and field-driven parity transition as ZBCs~\cite{lee2014spin,jellinggaard2016tuning}. Associated $\zeropi$ transitions were detected as a supercurrent reversals when the S-QD-S junction was embedded in a SQUID \cite{van2006supercurrent,Delagrange2016,li2019zeeman,razmadze2020quantumdot}. In the present study, we spectroscopically interrogate the ABS spectrum of a S-QD-S junction, which reveals the ability for these parameters to work in concert with a superconducting phase difference to cause parity transitions.

\section{Device}
\begin{figure}[t]
	\centering{\includegraphics[width=1\columnwidth]{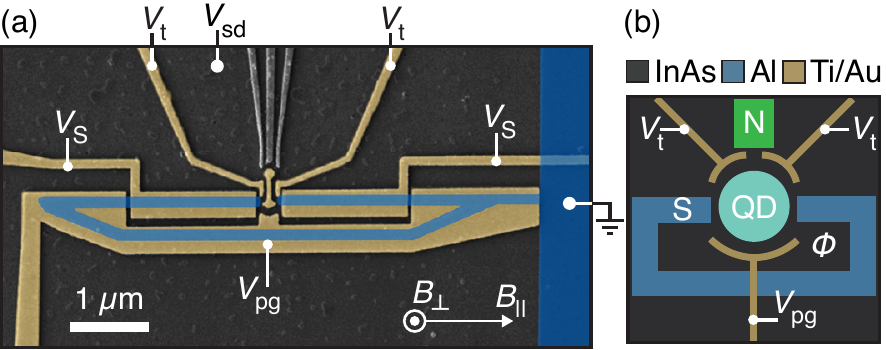}
	\caption{(a) False-color electron micrograph of an S-QD-S device, and (b) device schematic. Device consists of a loop (blue) of epitaxial Al, and Ti/Au electrostatic gates (yellow). The junction is formed by two Al leads defined by gate voltage $V_{\rm S}$ and confined into a quantum dot (QD) by gate voltage $\vpg$. The tunnel barriers to the normal (N) lead and superconducting leads are controlled by gate voltage $\vt$. ac+dc bias voltage $\vsd$ is applied to the normal lead with the superconducting loop  grounded. Magnetic field directions $\bpl$ and $\bperp$ are shown, where $\bperp$ is used to apply magnetic flux through the superconducting loop.  }
 	\label{fig1}}
\end{figure}

Devices were fabricated from an InAs two-dimensional electron gas (2DEG) heterostructure grown on InP with 8 nm of epitaxial Al deposited \emph{in-situ}. Details of the heterostructure stack and device fabrication are given in the Appendix. Previous measurements on similar material revealed near-unity transmission of the ABS in an S-N-S JJ~\cite{kjaergaard2017transparent,Nichele2020}.

We study two lithographically similar devices A and B. Figure \ref{fig1} shows a micrograph of the device A. The superconducting loop was selectively wet etched from the Al film. A 15 nm HfO$_2$ dielectric layer, grown by atomic layer deposition, was then deposited over the entire device. Ti/Au top-gates, patterned by electron-beam lithography, were then evaporated. The two S leads were defined by a negative gate voltage $V_{\rm S}$, forming a ballistic JJ of length 200 nm and connected by an Al loop to form a single-junction SQUID. The QD between the two S leads was defined and controlled with a negative voltage $\vpg$.  Typical QD charging energies varied between $U$ = 0.7 to 1 meV, giving $U/\Delta \sim 4 $ (see Supplementary Fig.~\ref{SI3}). Tunnel barriers to the N and S leads were controlled by gate voltage $\vt$ applied to both barriers, providing tunneling spectroscopy of the S-QD-S junction.  A voltage bias $\vsd$ consisting of ac and dc components was applied to the normal semiconductor (N) lead and the resulting current $I$ and four-terminal voltage $V_{\rm 4T}$ was measured using conventional lock-in techniques with the S loop grounded. The in-plane magnetic fields $\bpl$ and perpendicular field $\bperp$ were applied using a three-axis vector magnet. The superconducting phase difference $\varphi$ across the junction was controlled by threading magnetic flux through the S loop, which has area of 1.8 $\mu\rm{m}^2$, so that $\sim1.2$~mT corresponds to one flux quantum, $\Phi_{0}= h/2e=2.07$~mT$\,\mu$m$^{2}$.

\section{Model}

\begin{figure}[b]
	\centering{\includegraphics[width=1\columnwidth]{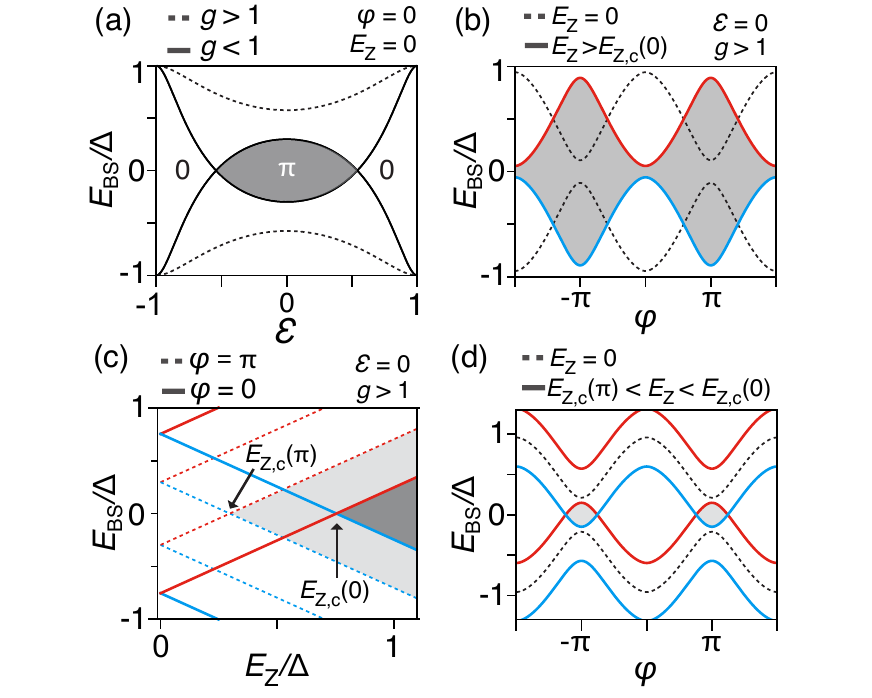}
	\caption{Subgap excitation spectrum of an S-QD-S Josephson junction (JJ) as a function of detuning, $\varepsilon$, centered on odd occupancy at $\varepsilon = 0$. (a) Bound-state energy $\Ebs(\varepsilon)$ normalized by gap, $\Delta$ for strong  ($g>1$, dashed) and weak ($g<1$, solid) coupling to the superconductors. Grey shaded region indicates odd-parity ground state. (b) $\Ebs$ as a function of superconducting phase difference $\varphi$ for an even (dashed) and odd (solid) parity ground state. (c) Dependence of  $\Ebs$ on Zeeman energy $\Ez = |g^*|\mu_{\rm B} \bpl$ for an even-parity ground state and for $\varphi=0, \pi$ (solid and dashed lines, respectively). A ground-state parity transition, from even to odd parity, occurs when $\Ebs = 0$ for a critical Zeeman energy $E_{\rm Z,c}$. (d) $\Ebs(\varphi)$ for an intermediate Zeeman energy ($E_{\rm Z,c}(\pi)<\Ez<E_{\rm Z,c}(0)$) where two zero-energy crossings occur near $\varphi = \pi$. Excitations are calculated based on a model introduced in Ref.~\cite{kirvsanskas2015yu} with an asymmetric lead coupling of $\theta \sim \pi/3$. Blue/red denote spin resolved subgap states. Black denotes spin degenerate states.}
	\label{fig2}}
\end{figure}

Before presenting experimental results, we first discuss the expected dependence of bound-state energies, $\Ebs$, of the QD on level detuning, $\varepsilon$, normalized Zeeman energy, $E_{\rm Z}/\Delta$, and phase difference, $\varphi$, across the dot, including parity and $\zeropi$ transitions (see Model Details for further information). Figure~\ref{fig2}(a) shows that for an odd-occupied QD, $\Ebs$ is lowered as $\varepsilon$ is tuned away from $\pm 1$, the QD charge degeneracy points. At the particle-hole symmetry point, $\varepsilon = 0$, the bound state energy depends on the parameter $g$, which describes the exchange coupling strength between the spin impurity and the S-lead as \cite{kirvsanskas2015yu,Zitko2016ZBA},
\begin{equation}
\Ebs= \Delta \frac{1-g^2}{1+g^2}\quad.
\end{equation}
For strong coupling ($g>1$, dashed curves in  Fig.~\ref{fig2}a), $\Ebs$ does not reach zero for any value $\varepsilon$, preserving an even-parity GS. Physically, this is a consequence of strong screening of the unpaired spin of the QD by a quasiparticle in the S leads, which together form a delocalized singlet~\cite{Zitko2016ZBA}. An increase in Coulomb interaction reduces $g$ that leads to a lowering of $\Ebs$. For $g< 1$, $\Ebs$ crosses zero energy, signalling a transition to an odd GS parity within the shaded region of Fig.~\ref{fig2}~\cite{Zitko2016ZBA}. 

Phase dependence of even and odd-parity GSs are shown in Fig.~\ref{fig2}b. Dashed curves show the phase dependence for the even-parity GS (a zero-junction, denoted 0-JJ), showing a $2\pi$ periodicity with $\Ebs$ minima at $\varphi = \pi$. For the odd-parity GS, the phase dependence acquires a $\pi$ shift, with energy minima at $\varphi = 0$ (solid curves), yielding a $\pi$ junction (denoted $\pi$-JJ). Notably, the $\zeropi$ transitions can be identified from this characteristic phase dependence of $\Ebs$.

Zeeman coupling, for instance from an in-plane magnetic field, can induce parity transitions by splitting an excited bound state doublet by the Zeeman energy, $E_{\rm Z} = |g^*|\mu_{\rm B} \bpl$, giving $\Ebs(\bpl) = \Ebs(0) \pm E_{\rm Z}/2 $~\cite{meden2019anderson,lee2014spin,He2020},  where $g^*$ is the effective g-factor and $\mu_{\rm B}$ is the  Bohr magneton~\cite{Zitko2016ZBA,yokoyama2014anomalous,vecino2003josephson,jellinggaard2016tuning}. In this scenario, at a critical Zeeman energy $E_{\rm Z,c}$, a zero-energy crossing results in a transition to an odd-parity GS, as shown in Fig.~\ref{fig2}c~\cite{lee2014spin,jellinggaard2016tuning}. Increasing the magnetic field further reopens a gap that stabilizes a magnetic doublet GS with a $\pi$-shifted phase dispersion~\cite{Wentzell2016}. Note that Zeeman-induced parity transitions occur at reduced $E_{\rm Z}$ for nonzero $\varphi$ (dashed curves in Fig.~\ref{fig2}c).

Figure \ref{fig2}d shows the phase dependence of an even-parity GS for Zeeman coupling in the range $E_{\rm Z,c}(\pi)<\Ez<E_{\rm Z,c}(0)$. In this intermediate range, $\Ebs$ is lowered such that a gap opens in the vicinity of $\varphi = \pi$, marked by two zero-energy crossings, indicating an odd-parity GS. The size of the resulting gap increases with increasing Zeeman energy until $\Ez$ reaches $E_{\rm Z,c}(0)$, resulting in a dispersion with two minima, one at $\varphi = 0$ and one at $\varphi = \pi$, denoted $0'$-JJ or $\pi'$-JJ, depending on which minimum is deeper \cite{rozhkov1999josephson,kirvsanskas2015yu}.

\section{Experiment}

\subsection{Gate voltage dependence}

\begin{figure}[t]
	\centering{\includegraphics[width=1\columnwidth]{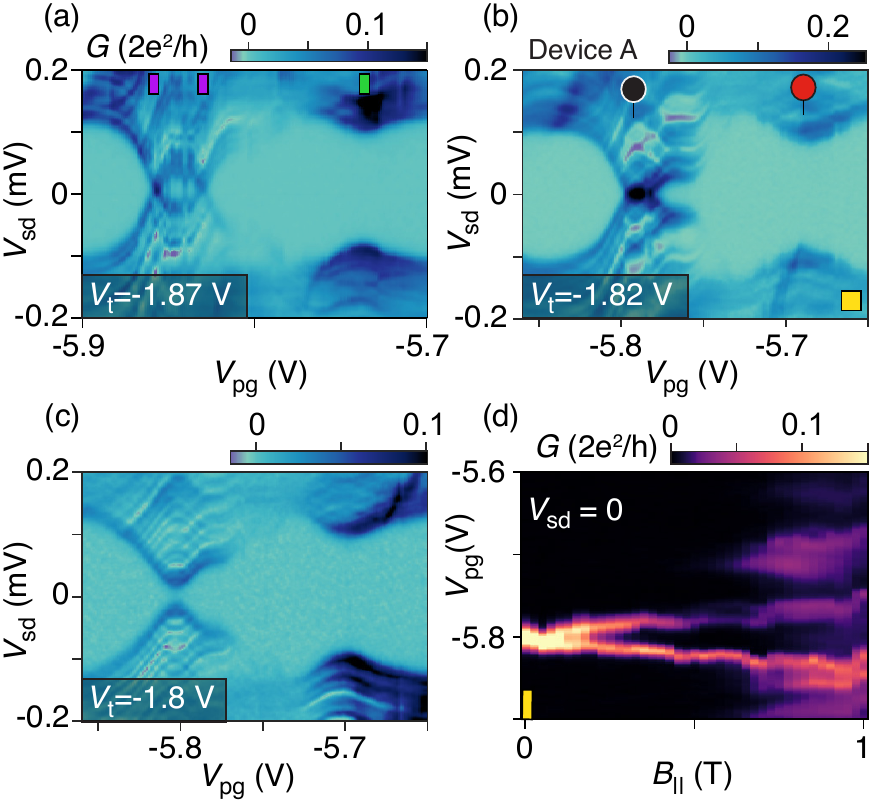}
	\caption{Parity transition induced by voltage $\vpg$ (device~A). Differential conductance $G$ as a function of bias  $\vsd$ and gate voltage $\vpg$ for tunnel-barrier gate voltage $\vt = $ -1.87~V (a), -1.82~V (b), and -1.8~V (c). (d) Zero-bias conductance, $G$, as a function of in-plane magnetic field $\bpl$ and $\vpg$ for $\vt = -1.82~V$. In panels a-c, $\bpl=\varphi=0$.}
	\label{fig3}}
\end{figure}

Tunneling spectroscopy of the S-QD-S junction was performed by creating a tunnel barrier to the normal lead ($\vt = -1.87$~V). In Fig.~\ref{fig3}a, the differential conductance $G = dI/dV_{\rm 4T}$ is shown as a function of bias voltage $\vsd$ and gate voltage $\vpg$ used to tune the occupancy of the QD. When $\vpg$ was varied, a gap of $\vsd \sim \pm 140~\mu$V was observed, along with two subgap features. The first feature occured at $\vpg \sim -5.74$~V (green marker in Fig.~\ref{fig3}a), where the gap is reduced to $\sim 100~\mu$V, indicating an odd QD occupancy with an even GS parity due to a strong coupling $g$ (see dashed line Fig \ref{fig2}a).

At the stronger subgap feature  in Fig.~\ref{fig3}a, around $\vpg \sim$ -5.85~V, the gap closed entirely, resulting in two ZBCs, indicated by purple markers. Increasing $\vt$ merged the ZBCs (black circle in Fig.~\ref{fig3}b), then removed the crossings entirely (Fig.~\ref{fig3}c). The sequence demonstrates a gate-voltage-induced GS parity transition of the type illustrated in Fig.~\ref{fig2}a, where $g$ is controlled by gate voltage $\vt$. The position in $\vpg$ of the two subgap features shift with $\vt$ due to cross coupling. Figure~\ref{fig3}d shows the dependence of the ZBC on $\bpl$ at the merging point, $\vt = -1.82$~V. The splitting is roughly linear in $\bpl$, yielding an effective g-factor $g^{*}\sim 5$.

%

\subsection{Phase dependence}
\begin{figure}
	\centering{\includegraphics[width=1\columnwidth]{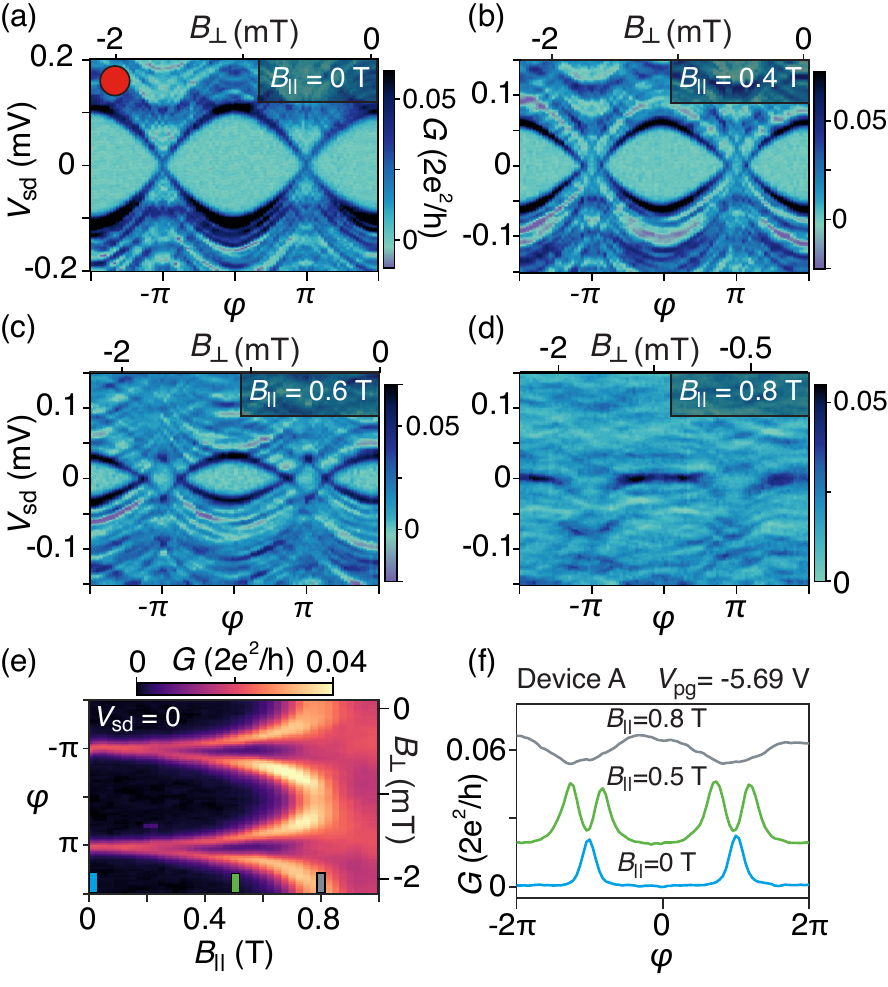}
	\caption{ Dependence of an even-parity ground state on superconducting phase difference $\varphi$ (device~A). (a) Differential conductance $G$ as a function of $\varphi$ and bias voltage $\vsd$ for in-plane magnetic field (a) $\bpl = 0$~T, (b) $\bpl = 0.4$~T, (c) $\bpl = 0.6$~T, (d) $\bpl = 0.8$~T. (e) Zero-bias $G$ as a function of $\varphi$  and $\bpl$ with vertically offset line-cuts shown in (f). The red marker indicates the $\vpg$ position in Fig.~\ref{fig3} for reference.}
	\label{fig4}}
\end{figure}

We next examine the phase dependence of the even-parity GS at the location of the red marker in Fig.~\ref{fig3}b ($\vpg = -5.69$~V, $\vt = -1.82$~V) by measuring the differential conductance $G$ as a function of $\varphi$ and $\vsd$, as shown in Fig.~\ref{fig4}a. Tuning $\varphi$ from 0 to $2\pi$ lowers $\Ebs$, eventually inducing a ZBC at $\varphi = \pi$. Applying an in-plane field $\bpl$ caused a gap to open in the vicinity of $\varphi = \pi$, as shown in Fig.~\ref{fig4}b. This gap increased with $\bpl$, while the gap at $\varphi = 0 $ decreased, as shown in Figs.~\ref{fig4}b-d. Although the induced superconducting gap is suppressed at $\bpl=0.8$~T (Fig.~\ref{fig4}d), it is evident that the minimum $\Ebs$ occurs at $\varphi = 0$. We interpret the gap opening at $\varphi = \pi$ as indicating $0'$-JJ behavior of the type illustrated in Fig.~\ref{fig2}d.

The position of the ZBC in both $\varphi$ and $\bpl$ is captured by measuring $G(\vsd$=0), as shown in Fig.~\ref{fig4}e. Increasing the field causes the crossing at $\varphi = \pi$ to split and move towards $\varphi = 0$, as shown in Fig.~\ref{fig4}f. The splitting, roughly linear at low fields, yields a g-factor $g^{*}\sim 5$, consistent with the value found from Fig.~\ref{fig3}d. Cuts in Fig.~\ref{fig4}f at $\bpl = 0$ and 0.8~T show minima shifted by $\pi$, indicating a $\zeropi$ transition driven by $\bpl$. 

Away from the feature marked by the red dot in Fig.~\ref{fig3}b, where the full gap is observed ($\vpg = -5.65$~V), ABSs do not cross zero-bias at $\varphi = \pi$ at zero magnetic field (see Supplemental Fig.~S.1).

\subsection{Magnetic field dependence}

\begin{figure}[b]
	\centering{\includegraphics[width=1\columnwidth]{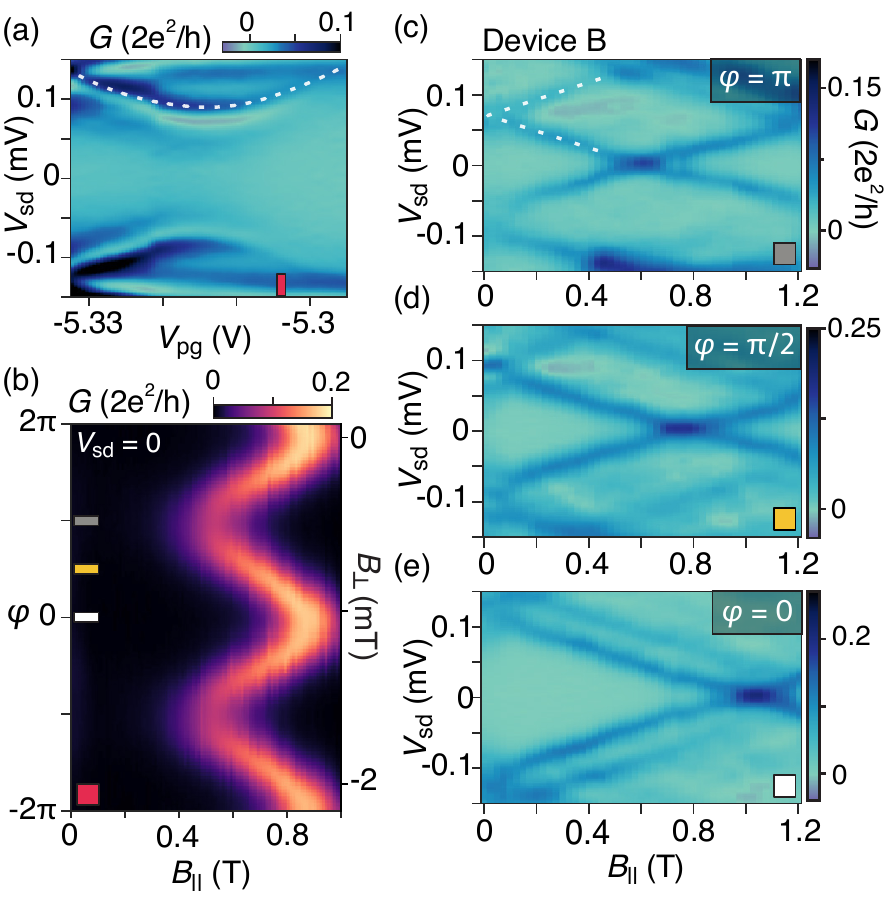}
	\caption{Magnetic field dependence of an even-parity ground state  (device~B). (a) Differential conductance $G$ as a function of gate voltage $\vpg$ and bias voltage $\vsd$ for $\varphi = \bpl = 0$. (b) Conductance $G$ at zero bias as a function of $\varphi$  and $\bpl$. $G$ as a function of bias $\vsd$ and in-plane field $\bpl$ for superconducting phase difference (c) $\varphi = \pi$, (d) $\pi/2$, (e) $0$. The plots in c-e are reconstructed from line cuts of Supplementary Fig.~\ref{SI4} for fixed $\varphi$ values. Dashed lines are guides to the eye.}
	\label{fig5}}
\end{figure}

We next investigate bound-state bias spectra as a continuous function of $\bpl$, rather than for the discrete values of $\bpl$ shown in Figs.~\ref{fig4}a-d. Focusing now on device B, Fig.~\ref{fig5}a shows a dip in $\Ebs$ as a function of gate voltage $\vpg$ without ZBCs, measured at $\bpl = 0$. This indicates an even-parity GS and a doublet ES throughout this range of $\vpg$.  Figure~\ref{fig5}b shows that a ZBC is first observed for $\bpl \sim 0.5$~T at a phase difference of $\pi$. Increasing $\bpl$ further causes the ZBC to split and merge at $\varphi = 0$, similar to Fig.~\ref{fig4}e. Comparing the phase dependence of the ZBC between $\bpl = 0.5$~T and $0.9$~T, it is clear that the position of the ZBC in $\varphi$ is $\pi$-shifted, indicating a magnetic field induced $\zeropi$ transition. We attribute the finite field needed to induce a parity transition in device B in comparison to device A to reflect a different coupling $g$ resulting from a different charging energy (see Fig.~\ref{figAp1}).

Figure \ref{fig5}c-e shows $\Ebs(\bpl)$ for fixed $\varphi$. At $\varphi = \pi$ (Fig.~\ref{fig5}c),  $\Ebs$ splits from its zero-field value of $\vsd = \pm 75~\mu{\rm V}$ moving linearly towards zero, crossing zero bias at $\bpl = 0.6$~T, indicating a field-driven GS parity transition. At $\varphi = \pi/2$ (Fig.~\ref{fig5}d) and $\varphi = 0$ (Fig.~\ref{fig5}e) the larger zero-field splittings push the zero-bias crossing point to larger $\bpl$. A g-factor $g^ *\sim 4$, extracted from the slope of the lower ES, is insensitive to $\varphi$. The dependence of the zero-bias crossing field on $\varphi$ is consistent with expectations in Fig.~\ref{fig2}c. Supplementary Fig.~\ref{SI4} shows spectroscopy of $\Ebs(\varphi)$ for fixed $\bpl$, displaying a continuous evolution of the $\zeropi$ transition.

\subsection{Odd-parity ground state}

\begin{figure}[t]
	\centering{\includegraphics[width=1\columnwidth]{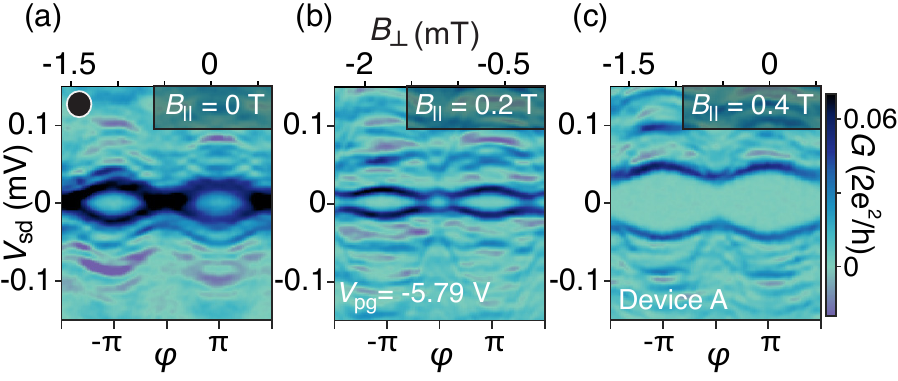}
	\caption{Phase dependence of the odd-parity ground state (device~A). Differential conductance $G$ as a function of superconducting phase difference $\varphi$ and bias $\vsd$ for (a) $\bpl = 0$, (b) $0.2 $~T, (c) $\bpl = 0.4$~T. Black marker indicates the $\vpg$ position in Fig. \ref{fig3} for reference.}
	\label{fig6}}
\end{figure}

The odd-parity transition indicated by the black marker in Fig.~\ref{fig3}b) is investigated in Fig.~\ref{fig6}. At zero field, $\Ebs(\varphi)$ is $\pi$-shifted compared to the even-parity case (Fig.~\ref{fig4}), with a minimum at $\varphi = 0$, indicating a $\pi$ junction. Near the odd-parity transition, $\Ebs(\varphi)$ shows a reduced dependence on $\varphi$, as reported previously~\cite{chang2013tunneling,pillet2010andreev}. Increasing $\bpl$ increases $\Ebs$, opening a gap for all $\varphi$ that increases with field while retaining the $\pi$ phase shift. This behaviour is consistent with theory (see Fig. \ref{fig2})~\cite{Wentzell2016}.


\subsection{Zero-bias crossings}

\begin{figure}[t]
\centering{
	\includegraphics[width=1\columnwidth]{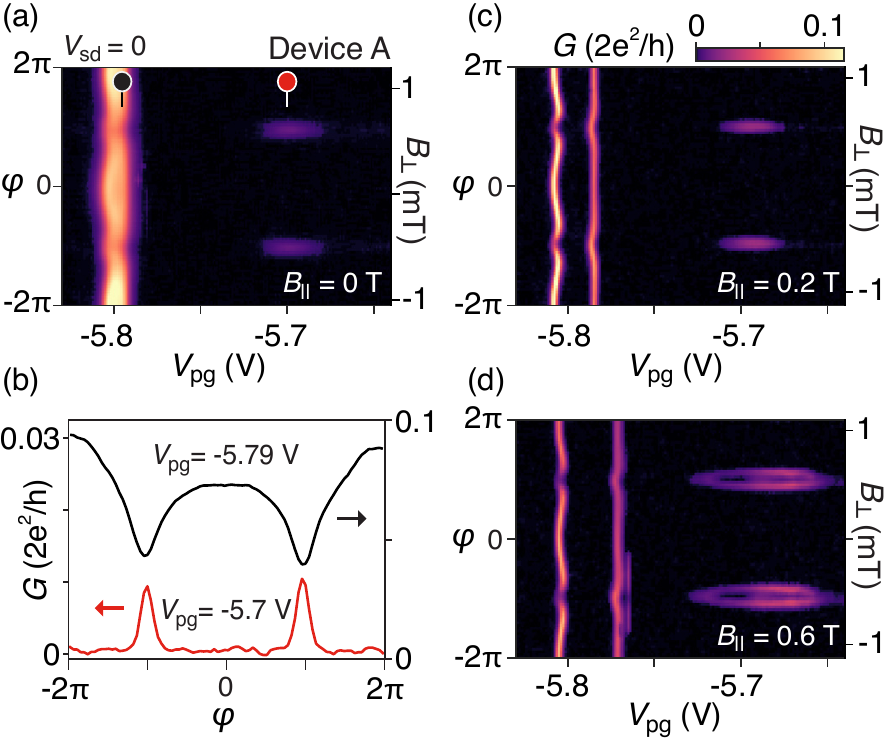}
	\caption{Evolution of zero-bias crossings (device A). (a) Differential conductance $G$ at zero bias as a function of  phase difference $\varphi$ and gate voltage $\vpg$ for (a) $\bpl = 0$. Red and black markers show positions of line cuts in (b). Note opposite behavior along the two line cuts, as discussed in the text. (b) Line-cuts of (a) for an even-parity (red) and odd-parity (black) ground states. (c) Same as (a) for $\bpl = 0.2$~T, (d) Same as (a) for $\bpl = 0.6$~T.}
	\label{fig7}}
\end{figure}

The contribution of in-plane magnetic field, phase difference, and gate voltage on GS parity and $\zeropi$ transitions is identified by measuring $G(\vpg,\varphi)$ at zero-bias in device A (see Fig.~\ref{fig7}).  This allows for the phase dependence of the ZBCs to be highlighted at specific $\vpg$ values. Figure \ref{fig7}a shows two distinct values of $\vpg$ where ZBCs occur at $\bpl = 0$ (see red and black markers). 

At $\vpg = -5.7$~V, ZBCs are observed at $\varphi = \pi$, marking the position of the even-parity GS investigated in Fig.~\ref{fig4}. We interpret the limited range of this ZBC in $\vpg$ to reflect the energy dependence of $\Ebs(\varepsilon)$ illustrated in Fig \ref{fig2}a, with the red marker signifying $\varepsilon = 0$. For increasing magnetic field the ZBC at $\varphi = \pi$ splits in both $\varphi$ and $\vpg$, stabilizing an odd-parity GS (see Fig.~\ref{fig7}d).

For $\vpg = -5.8$~V, a bright vertical band is observed that indicates the odd-parity GS examined in Fig.~\ref{fig6}. In Fig.~\ref{fig7}b the phase dependence of the two GS locations is compared that reveals a $\pi$-shifted $\Ebs(\varphi)$ dependence for the odd-parity GS. Increasing $\bpl = 0.2$~T causes the odd occupancy ZBC to split in $\vpg$ while retaining a $\pi$-shifted phase dependence with $g^ *\sim 5$. The results of Fig.~\ref{fig7} reveals how the combination of gate-voltage, magnetic field, and phase difference can control subgap excitations of the system and induce GS parity transitions.

\section{Conclusion}

To summarize, we have measured the subgap spectrum of an S-QD-S Josephson junction under the influence of gate voltage, in-plane magnetic field, and superconducting phase difference. We found that odd QD occupancies were not always accompanied by parity transitions or $\pi$-shifted Andreev spectrum. However, by controlling either the coupling, magnetic field, or phase difference, subgap excitations could be lowered to zero bias, inducing a parity transition. Furthermore, we showed that by applying a finite phase difference across the junction, parity transitions can occur at lower magnetic fields.

These results may have important implications for semiconductor based superconducting qubits~\cite{casparis2018superconducting}, which recently showed that an unintentional QD resonance resulted in a suppressed charge dispersion~\cite{Kringhoej2020,Bargerbos2020}. We demonstrate that highly tunable QDs can be intentionally placed in the weak link that may enable an alternative mechanism for charge noise suppression while retaining large qubit anharmonicity. Moreover, these results introduce novel means of manipulating the spin of the ABS that could be used for controlling Andreev qubits~\cite{Hays2018,Tosi2019}.

Our results demonstrate both the high material quality and device design flexibility offered by the InAs-Al heterostructure material platform. The S-QD-S device design studied here is a promising candidate for investigating the hybridization of a QD with Majorana zero modes in the pursuit of parity readout of a topological qubit~\cite{Karzig2017}. 

\bibliography{PiBib.bib}

\noindent \textbf{Acknowledgments} This work was supported by Microsoft Corporation, the Danish National Research Foundation, and the Villum Foundation. We thank Karsten Flensberg, Jens Paaske, and Jens Schulenborg for useful discussions.

\section*{Appendix}

\subsection{Wafer structure}
The wafers used for fabricating the devices were grown by molecular beam epitaxy. The material stack consists of an InP substrate with a 100-nm-thick $\rm{In_{0.52}Al_{0.48}As}$ lattice matched buffer, a 1-$\mathrm{\mu m}$-thick step-graded buffer realized with alloy steps from $\rm{In_{0.52}Al_{0.48}As}$ to $\rm{In_{0.89}Al_{0.11}As}$ (20 steps, 50 nm/step), a $58~\rm{nm}$ $\rm{In_{0.82}Al_{0.18}As}$ layer, a $4~\rm{nm}$ $\rm{In_{0.75}Ga_{0.25}As}$ bottom barrier, a $7~\rm{nm}$ InAs quantum well, a $10~\rm{nm}$ $\rm{In_{0.75}Ga_{0.25}As}$ top barrier, two monolayers of GaAs and a $7~\rm{nm}$ film of epitaxial Al deposited \textit{in-situ} without breaking the MBE chamber vacuum. 

Hall bar device geometries (where the Al was removed) were used to characterize the two-dimensional electron gas and revealed an electron mobility peak $\mu=43,000~\rm{cm^2V^{-1}s^{-1}}$ for an electron density $n=8\times10^{11}~\rm{cm^{-2}}$, corresponding to an electron mean free path of $l_{\rm e} \sim 600$ nm.

\subsection{Fabrication Details}
Devices were fabricated using standard electron beam lithography  and wet etching techniques. The devices were electrically isolated using a two-step mesa etch by first removing the top Al film with Al etchant Transene D, and then a deep $\sim 300$~nm III-V chemical wet etch $\rm{H_2 O:C_6 H_8 O_7:H_3 PO_4:H_2 O_2}$ (220:55:3:3). In a following lithography step, the Al film on the mesa was selectively etched into a SQUID with Al etchant Transene D at $50^{\circ}\mathrm{C}$. A $15~\rm{nm}$ thick layer of insulating $\rm{HfO_2}$ was grown over the entire sample by atomic layer deposition at a temperature of $90^{\circ}\mathrm{C}$.  Finally, top gates of Ti/Au (5/25$~\rm{nm}$) were deposited by electron beam evaporation and connected to bonding pads with leads of Ti/Au (5/300$~\rm{nm}$).

\subsection{Measurement Details}
Electrical measurements were performed in a dilution refrigerator at a base temperature of $20~\rm{mK}$. Using conventional lock-in techniques at 166 Hz, an ac excitation voltage of 3~$\mu \rm{V}$ and a variable dc bias voltage $\vsd$ was applied to the normal lead ohmic as shown in Fig.~\ref{fig1}. The resulting current across the device was recorded by grounding the superconducting loop ohmic via a low-impedance current-to-voltage converter, and the four terminal voltage was measured by an ac voltage amplifier with an input impedance of $500~\rm{M\Omega}$. 

\subsection{Model Details}

The energy of the subgap excitations in an S-QD-S system was theoretically examined under the influence of QD level detuning $\varepsilon$, coupling to the superconductor $g$, magnetic field $\bpl$, and superconducting phase difference $\varphi$ with the model proposed by  Kir{\v{s}}anskas, G. \textit{et al.}~\cite{kirvsanskas2015yu}. This is an Anderson-type model describing a single Coulomb blockaded QD level that is coupled to two S leads with superconducting gaps $\Delta \exp(\pm i\varphi/2)$ in the limit of $U\gg\Delta$. The excitations in this model are Yu-Shiba-Rosinov states resulting from spinful odd QD occupancies.

In Fig.~\ref{fig2} we examine the energy of bound-state excitations $\Ebs$ in an S-QD-S JJ. The bound-state energies are calculated from~\cite{kirvsanskas2015yu},
\begin{equation}
\begin{split}
\MoveEqLeft
E_{\pm, \sigma} = \frac{1}{2}\sigma \Ez - \frac{\sigma c_{\pm}\Delta}{\sqrt{(1+u)^2+4g^2}}\Big[(1+u)(1+\chi u)  \\
&+2g^2\pm 2g\sqrt{g^2+u(1-\chi)(1+\chi u)}\Big]^{1/2}
\end{split}
\label{eq.YSR}
\end{equation}
where the following shorthand notation is used,
\begin{equation}
\begin{split}
\MoveEqLeft
\chi = 1-\sin^2(2\theta)\sin^2(\phi/2), \quad u=w^2-g^2, \quad  c_+ = 1\\
&c_- = \rm{sign}(1+\chi u),  \quad \tan(\theta) = t_R/t_L \quad .
\end{split}
\end{equation}
The exchange scattering amplitude $g$ and the potential scattering amplitude $w$ both depend on the position of the QD level detuning $\varepsilon$. The spin of the bound-states $\sigma=\pm1$ is either aligned or anti-aligned with respect to the spin of the QD. An angle $\theta$ is introduced to account for an asymmetry between the left and right tunnel barriers ($t_{R/L}$) to the superconducting leads, where $\theta = \pi/4$ represents a symmetrical coupling.  In Ref. \cite{kirvsanskas2015yu} these bound-state energies are used to calculate the conductance with a weakly coupled normal lead (similar setup as in Fig.~\ref{fig1}b), where a good agreement between the simulated conductance and the subgap spectra shown in Fig.~\ref{fig2} is found.

\begin{figure}[h]
	\centering{\includegraphics[width=0.9\columnwidth]{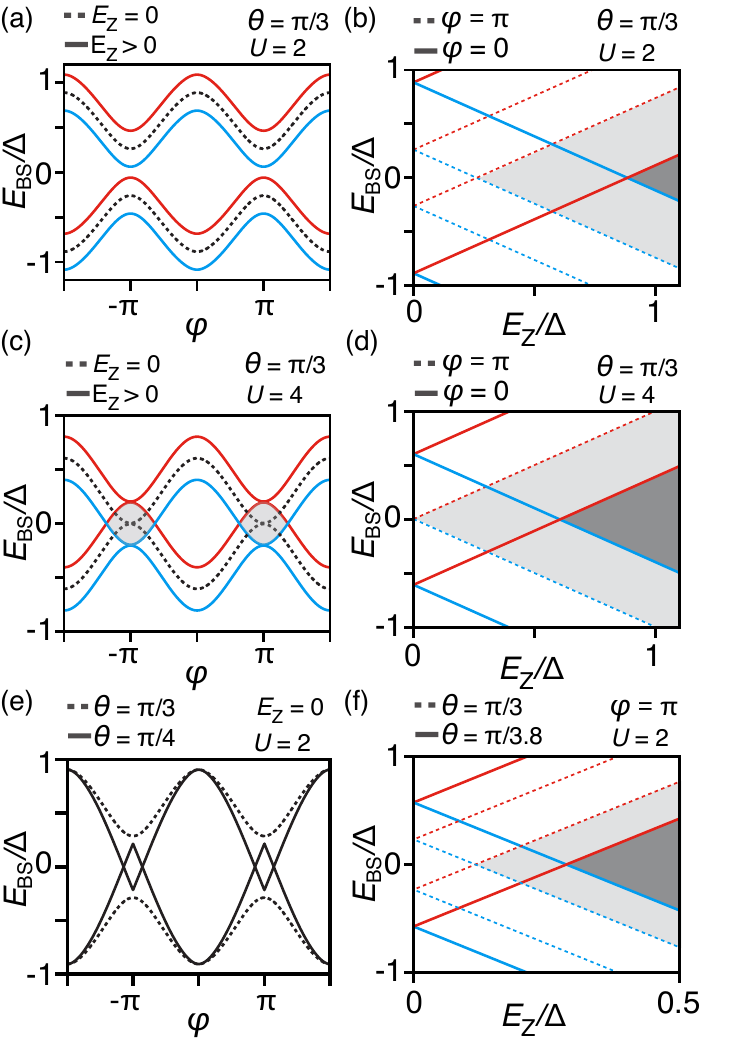}
	\caption{Dependence of bound-state spectra on superconducting phase difference $\varphi$ and Zeeman energy $\Ez$  for (a,b) charging energy $U$ = 2 and (c,d) $U$ = 4 for asymmetric coupling to the superconducting leads $\theta = \pi/3$. (e,f) Dependence of asymmetric coupling $\varphi$ on phase dispersion and Zeeman energy of the bound-state excitations.}
	\label{figAp1}}
\end{figure}

In Fig. \ref{figAp1}(a-d) the dependence of a varying charging energy $U$ is shown. By decreasing the charging energy, a gap at $\varphi = \pi$ opens up due to an increased $g$. This shifts the critical Zeeman energy to higher fields (see Fig. \ref{figAp1} b, d). Figure~\ref{figAp1}(e,f) shows the effect of coupling asymmetry on the phase dispersion. Asymmetric left/right coupling can open a gap at $\varphi=\pi$, which can be closed by symmetrizing the coupling or applying a magnetic field.

In the model of Kir{\v{s}}anskas \textit{et al.}~\cite{kirvsanskas2015yu}, a polarized spin approximation on the QD is employed to derive Eq. \ref{eq.YSR}. Therefore, the Zeeman energy does not influence the QD but induces spin splitting in the superconducting leads. Experimentally we interpret the observed magnetic field dependence to reflect Zeeman splitting of the doublet ground states as discussed theoretically in Refs.~\cite{jellinggaard2016tuning,Zitko2016ZBA} and experimentally in Ref.~\cite{lee2014spin}. Experimentally it is challenging to differentiate between the two Zeeman splitting mechanisms since they contribute different g-factor values as discussed in Ref.~\cite{Zitko2017}. We therefore assume an effective g-factor $g^*$ that accounts for a contribution from both mechanisms.

\clearpage

\setcounter{figure}{0}
\renewcommand{\thefigure}{S.\arabic{figure}}
\renewcommand{\theHfigure}{Supplement.\thefigure}

\onecolumngrid
\clearpage
\onecolumngrid

\section{Supplementary Information}

\begin{figure*}[h]
	\includegraphics[width=1\columnwidth]{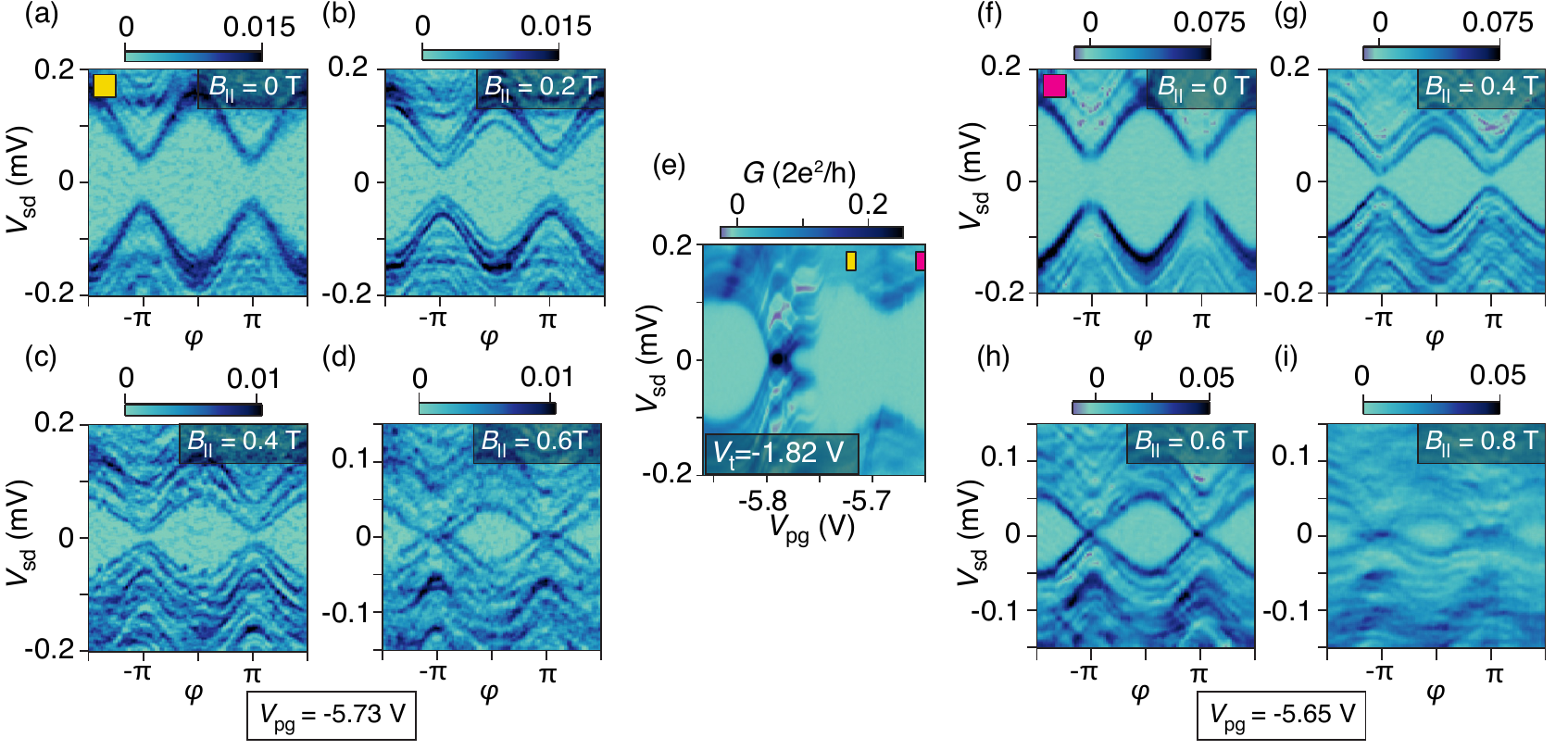}
	\caption{Dependence of a even-parity ground state on a superconducting phase difference $\varphi$ for intermediate values of $\vpg$ in device~A. (a-d) Differential conductance $G$ as a function of $\varphi$ and bias voltage $\vsd$ for magnetic field $\bpl = 0$~T (a), 0.2~T (b), 0.4~T (c), 0.6~T (d) for $\vpg=-5.73$~V. (e) $G$ as a function of $\vsd$ and $\vpg$ for tunnel barrier gate voltage $\vt = $ -1.82~V. (f-i) $G$ as a function of $\varphi$ and $\vsd$ for $\bpl = 0$~T (f), 0.4~T (g), 0.6~T (h), 0.8~T (i) for $\vpg=-5.65$~V.
	}
	\label{SI1}
\end{figure*}

\begin{figure*}[h]
	\includegraphics[width=1\columnwidth]{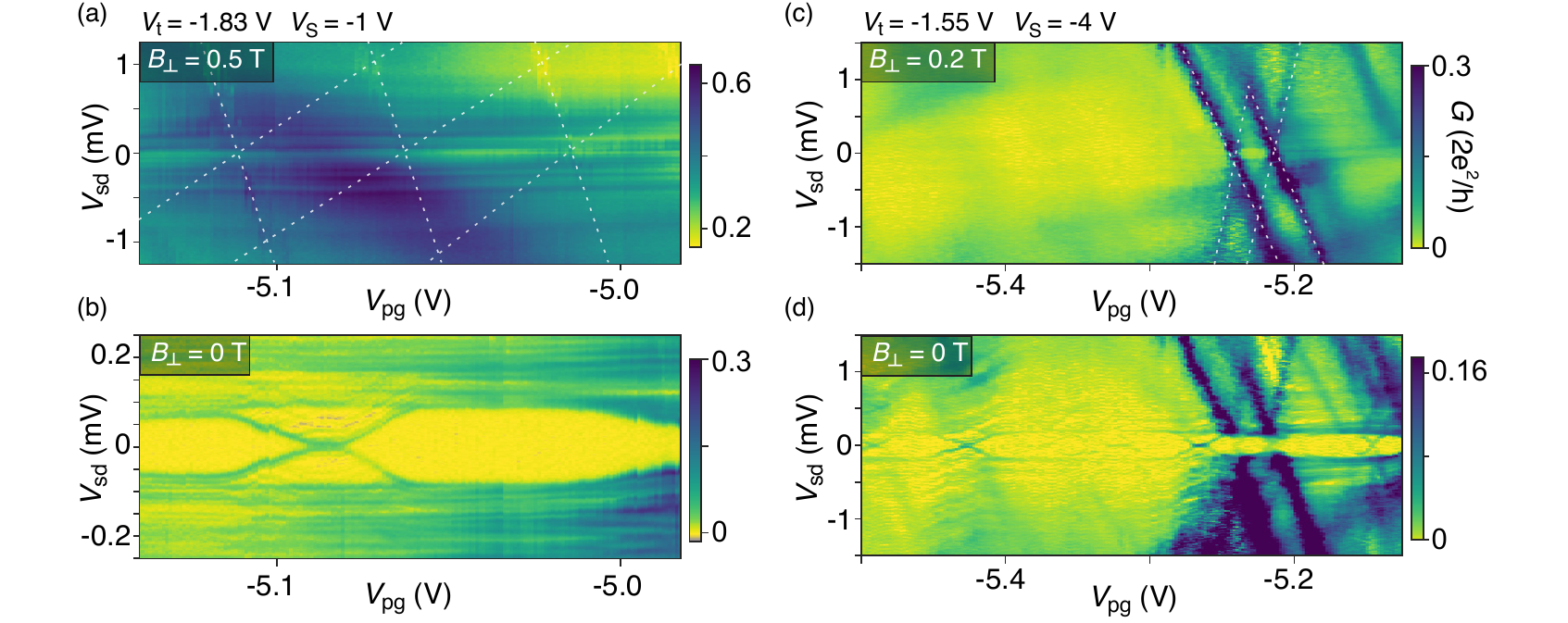}
	\caption{Coulomb blockade in device~A. (a-d) Differential conductance $G$ as a function of bias voltage $\vsd$ and $\vpg$ in the normal state (a,c) and in the superconducting state (b,d). Panels (a) and (b) are measured in a similar gate configuration shown in Fig.~\ref{fig3}b. Panels (c,d) are measured with more negative $V_{\rm S}$ gate voltages to allow for clearer Coulomb blockade features.
	}
	\label{SI3}
\end{figure*}

\begin{figure*}[h]
	\includegraphics[width=1\columnwidth]{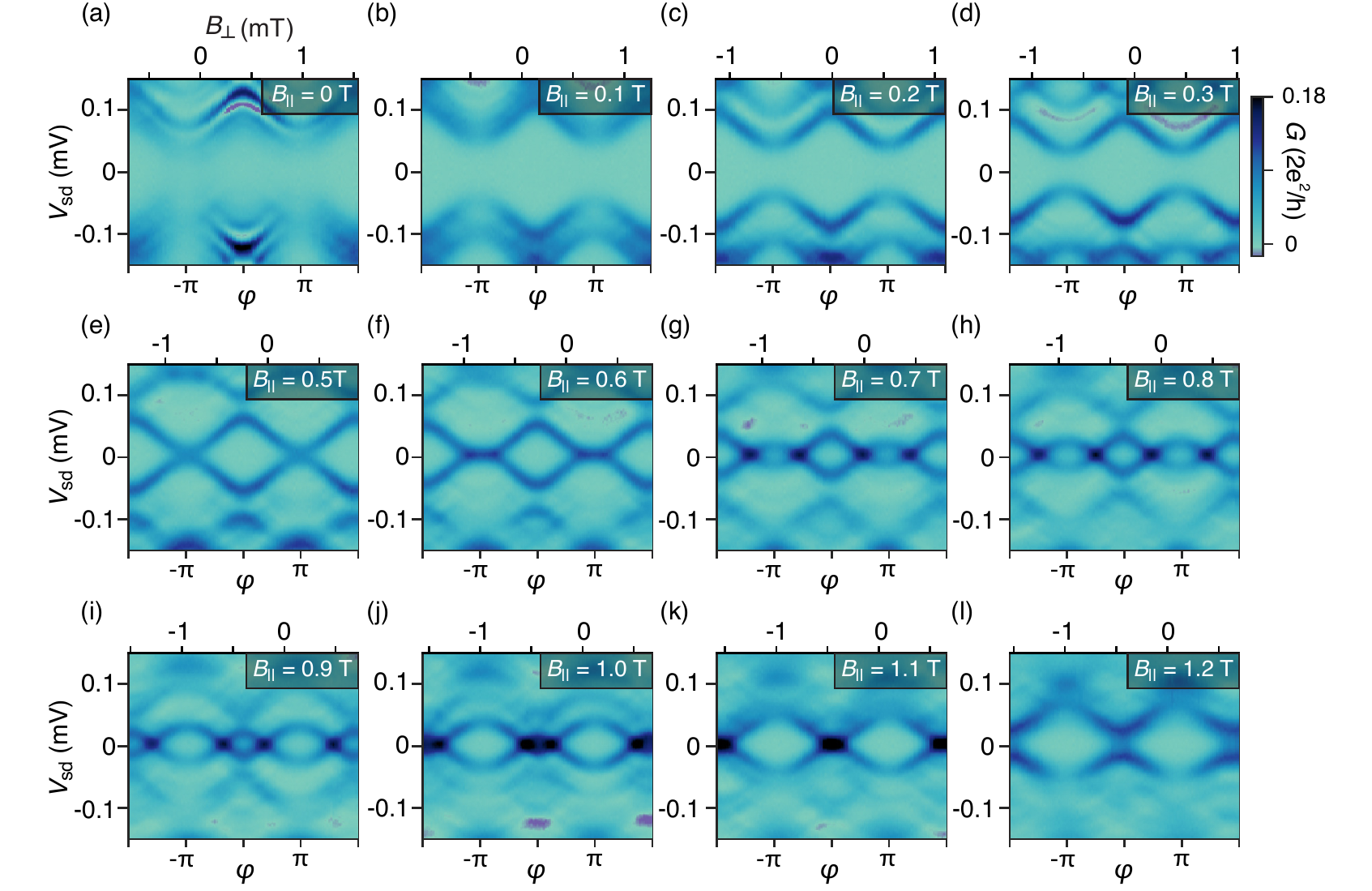}
	\caption{Evolution of phase dispersion in magnetic field for device B. (a-l) Differential conductance $G$ as a function of $\varphi$ and bias voltage $\vsd$ for increasing magnetic field $\bpl$.
	}
	\label{SI4}
\end{figure*}

\end{document}